\documentclass[prd,reprint,nofootinbib,notitlepage,aps,tightenlines,preprintnumbers,amsmath,amssymb,showpacs,superscriptaddress]{revtex4-1}

\usepackage[hyperindex,breaklinks]{hyperref}
\usepackage{graphicx,epstopdf,amsmath,amsfonts,amssymb,color}

\renewcommand{\sec}{\ensuremath{\mathrm{s}}}

\newcommand{\kg}{\ensuremath{\mathrm{kg}}}

\newcommand{\cm}{\ensuremath{\mathrm{cm}}}

\newcommand{\eV}{\ensuremath{\mathrm{eV}}}
\newcommand{\keV}{\ensuremath{\mathrm{keV}}}

\newcommand{\GeV}{\ensuremath{\mathrm{GeV}}}

\renewcommand{\day}{\ensuremath {\mathrm{day}}}

\DeclareMathOperator{\real}{Re}
\DeclareMathOperator{\imag}{Im}

\begin{document}

\preprint{CALT-TH 2014-173}

\title{Direct Detection Constraints on Dark Photon Dark Matter}

\author{Haipeng An}
\affiliation{Walter Burke Institute for Theoretical Physics,
California Institute of Technology, Pasadena, CA 91125}

\author{Maxim Pospelov}
\affiliation{Department of Physics and Astronomy, University of Victoria, 
Victoria, BC V8P 5C2, Canada}
\affiliation{Perimeter Institute for Theoretical Physics, Waterloo, ON N2J 2W9, 
Canada}

\author{Josef Pradler}
\affiliation{Institute of High Energy Physics, Austrian Academy of
  Sciences, Nikolsdorfergasse 18, 1050 Vienna, Austria}

\author{Adam Ritz}
\affiliation{Department of Physics and Astronomy, University of Victoria, 
Victoria, BC V8P 5C2, Canada}

\date{December 2014}

\begin{abstract}
Dark matter detectors built primarily to probe elastic scattering of WIMPs on nuclei are also precise probes of light, weakly coupled, particles that may be absorbed by the detector material.  In this paper, we derive constraints on the minimal model of dark matter comprised of long-lived vector states $V$ (dark photons) in the $0.01-100$ keV mass range.  The absence of an ionization signal in direct detection experiments such as XENON10 and XENON100 places a very strong constraint on the dark photon mixing angle, down to ${\cal O}(10^{-15})$, assuming that dark photons comprise the dominant fraction of dark matter. This sensitivity to dark photon dark matter exceeds the indirect bounds derived from stellar energy loss considerations over a significant fraction of the available mass range. We also revisit indirect constraints from $V\to 3\gamma$ decay and show that limits from modifications to the cosmological ionization history are comparable to the updated limits from the diffuse $\gamma$-ray flux.
\end{abstract}
\maketitle

\section{Introduction}

The Standard Model of particle physics (SM) is known to be incomplete,
in that it needs to be augmented to include the effects of neutrino
mass. Furthermore, cosmology and astrophysics provide another strong
motivation to extend the SM, through the need for dark matter (DM).
Evidence ranging in distance and time scales from the horizon during
decoupling of the cosmic microwave background (CMB) to sub-galactic
distances points to the existence of `missing mass' in the form of cold,
non-baryonic DM. The particle (or, more generally, field theoretic)
identity of dark matter remains a mystery -- one that has occupied the
physics community for many years.

While the `theory-space' for DM remains enormous, several model
classes can be broadly identified.  Should new physics exist at or
near the electroweak scale, a weakly interacting massive particle
(WIMP) becomes a viable option. The WIMP paradigm assumes the
existence of a relatively heavy particle (typically with a mass in the
GeV to TeV range) having sizeable couplings to the SM.  The
self-annihilation into the SM regulates the WIMP cosmic abundance
according to thermal freeze-out, and the observed relic density
requires a weak-scale annihilation rate. The simplest models of this
type also predict a significant scattering rate for WIMPs in the
galactic halo on nuclei, when up to $100~{\rm keV}$ of WIMP kinetic
energy can be transferred to atoms, offering a variety of pathways for
detection. Direct detection, as it has became known, is a rapidly
growing field, with significant gains in sensitivity achieved in the
last two decades, and with a clear path forward
\cite{Cushman:2013zza}.

Alternatively, DM could be in the form of super-weakly interacting
particles, with a negligible abundance in the early Universe, and 
generated through a sub-Hubble thermal leakage rate (also known as
the `freeze-in' process). Dark matter of this type is harder to detect
directly, as the couplings to the SM are usually smaller than those of
WIMPs by many orders of magnitude. Metastability of such states offers
a pathway for the indirect detection of photons in the decay products,
as is the case for metastable neutrino-like particles in the
$O(10~\rm keV)$ mass range (see, {\em e.g.}
\cite{Abazajian:2006yn}). It was also pointed out in
\cite{Pospelov:2008jk} that WIMP direct detection experiments are
sensitive to bosonic DM particles with couplings of $O(10^{-10})$
or below, that could be called super-WIMPs (referring to the
`super-weak' strength of their SM interactions).

Finally, a completely different and independent class of models for
dark matter involves light bosonic fields with an abundance generated
via the vacuum misalignment mechanism
\cite{Preskill:1982cy,Abbott:1982af,Dine:1982ah}. In this class of models, DM
particles emerge from a cold condensate-like state with very large
particle occupation numbers, which can be well described by a
classical field configuration. The mass and initial amplitude of the
DM field defines its present energy density. The most prominent
example in this class, the QCD axion, does have a non-vanishing
interaction with SM fields, although other forms of `super-cold'
DM do not necessarily imply any significant coupling.  While axion
dark matter has been the focus of many experimental searches and
proposals~\cite{Jaeckel:2010ni}, other forms of super-cold dark
matter have received comparatively less attention (see {\em e.g.}~\cite{Matos:2008ag,Piazza:2010ye,Nelson:2011sf,Horns:2012jf}).
In the course of these latter investigations, and subsequent work,
several experimental strategies for detecting such dark matter
scenarios have been suggested
\cite{Jaeckel:2013sqa,Arias:2014ela,Chaudhuri:2014dla}.

Regarding the latter class of models, it is also possible to generate
a dark matter abundance not only from a pre-existing condensate
(vacuum misalignment) but gravitationally, during inflation, through
perturbations in the field that carry finite wave number
$k$~\cite{Mukhanov:1990me}. Recent work~\cite{Graham:2015rva}
investigates this possibility for vector particles, reaching the
conclusion that such mechanism avoids large-scale isocurvature
constraints from CMB observations, and allows light vectors to be
generated in sufficient abundance as viable dark matter candidates.

\begin{figure*}
\centering
\includegraphics[width=0.55\textwidth]{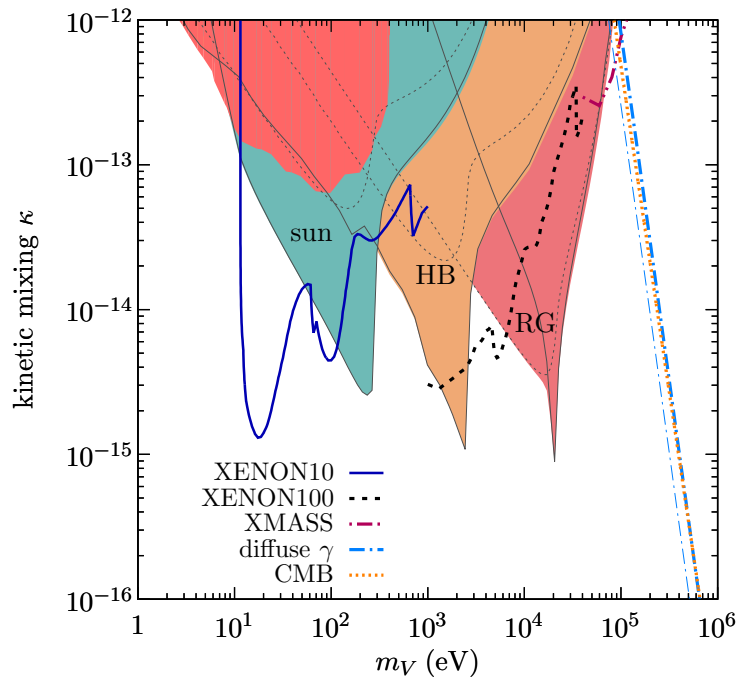}%
\caption{\footnotesize A summary of constraints on the dark photon
  kinetic mixing parameter $\kappa$ as a function of vector mass $m_V$
  (see Secs.~\ref{sec:dark-photon-dark} and \ref{sec:dm-absorpt-sign}
  for the details). The thick lines exclude the region above for dark
  photons with dark matter relic density. The solid (dashed) line is
  from XENON10 (XENON100); the limit from XMASS is taken
  from~\cite{Abe:2014zcd}. The dash-dotted lines show our newly
  derived constraints on the diffuse $\gamma$-ray flux from
  $V\to 3\gamma$ decays, assuming that decays contribute 100\% (thick
  line) or 10\% (thin line) to the observed flux. The thick dotted
  line is the corresponding constraint from CMB energy
  injection. Shaded regions depict (previously considered)
  astrophysical constraints that are independent of the dark photon
  relic density. The limits from anomalous energy loss in the sun
  (sun), horizontal branch stars (HB), and red giant stars (RG) are
  labeled. The shaded region that is mostly inside the solar
  constraint is the XENON10 limit derived from the solar flux~\cite{An:2013yua}.}
\label{fig:summary}
\end{figure*}

In this paper, we consider `dark photon dark matter' generated through
inflationary perturbations, or possibly other non-thermal mechanisms.
While existing proposals to detect dark photons address the range of
masses below ${\cal O}$(meV), we will investigate the sensitivity of
existing WIMP-search experiments to dark photon dark matter with mass
in the 10 eV - 100 keV window.  As we will show, the coupling constant
of the dark photon to electrons, $e\kappa$, can be probed to
exquisitely low values, down to mixing angles as low as
$\kappa \sim O(10^{-15})$.  Furthermore, sensitivity to this mixing
could be improved with careful analysis of the `ionization-only'
signal available to a variety of DM experiments. The sensitivity of
liquid xenon experiments to vector particles has already been explored
in~\cite{Arisaka:2012pb} and many experiments have already reported
relevant analyses
\cite{Bernabei:2005ca,Aalseth:2008rx,Ahmed:2009rh,Horvat:2011jh,Armengaud:2013rta,Abgrall:2014bka,Aprile:2014eoa,Abe:2014zcd}.
While we concentrate on the Stuckelberg-type mass for the vector
field, our treatment of direct detection of $V$ will equally apply to
the Higgsed version of the model. Moreover, the existence of a Higgs
field charged under $U(1)'$ opens up additional possibilities for
achieving the required cosmological abundance of~$V$.

The rest of this work is organized as follows. In
Sec.~\ref{sec:dark-photon-dark} we introduce the dark photon model
in some more detail, describe existing constraints, and reconsider
indirect limits. In Sec.~\ref{sec:dm-absorpt-sign} we compile the
relevant formul\ae ~for direct detection, confront the model with
existing direct detection results and derive constraints on the mixing
angle $\kappa$. The results are summarized in Fig.~\ref{fig:summary},
which shows the new direct detection limits in comparison to various
astrophysical constraints. In Sec.~\ref{sec:disc-concl}, we provide a general
discussion of super-weakly coupled DM, and possible improvements in
sensitivity to (sub-)keV-scale DM particles.

\section{Dark photon dark matter} 
\label{sec:dark-photon-dark}

It has been well-known since 1980s that the SM allows for a natural UV-complete
extension by a new massive or massless $U(1)'$ field, coupled to the
SM hypercharge $U(1)$ via the kinetic mixing term
\cite{Holdom:1985ag}. Below the electroweak scale, the effective
kinetic mixing of strength $\kappa$ between the dark photon ($V$) and
photon ($A$) with respective field strengths $V_{\mu\nu}$ and
$F_{\mu\nu}$ is the most relevant,
\begin{align}
  \label{eq:L}
  \mathcal{L} = -\frac{1}{4} F_{\mu\nu}^2-\frac{1}{4} V_{\mu\nu}^2 -
  \frac{\kappa}{2} F_{\mu\nu}V^{\mu\nu} + \frac{m_V^2}{2} V_{\mu}V^{\mu}
  + e J_{\mathrm{em}}^{\mu} A_{\mu},
\end{align}
where $ J_{\mathrm{em}}^{\mu} $ is the electromagnetic current and
$m_V$ is the dark photon mass.
This model has been under significant scrutiny over the last few
years, as the minimal realization of one the few UV-complete extensions of the SM
(portals) that allows for the existence of light weakly coupled particles
\cite{Essig:2013lka}.  For simplicity, we will consider the
St\"uckelberg version of this vector portal, in which $m_V$ can be
added by hand, rather than being induced via the Higgs mechanism.

\subsection{Cosmological abundance}
\label{sec:cosm-abund}

Light vector particles with $m_V < 2 m_e$ have multiple
contributions to their cosmological abundance, such as (a) production
through scattering or annihilation, $\gamma e^{\pm}\to V e^{\pm}$ and
$e^+e^-\to V\gamma$, possibly with sub-Hubble rates, (b) resonant
photon-dark photon conversion, or (c) production from an initial dark
photon condensate, as could be seeded by inflationary
perturbations. Notice that if mechanisms (a) and (b) are the only
sources that populate the DM, they are not going to be compatible with
cold dark matter when $m_V \lesssim$ keV.

For mechanism (a), naive dimensional analysis suggests a dark photon
interaction rate
$\Gamma_{\rm int} \sim \kappa^2 \alpha^2 n_e/s$, where $n_e$ is
the electron number density and $\sqrt{s}$ is the centre-of-mass
energy. At temperatures $T\gg m_e$, where the number density of charge
carriers is maximal, $n_e\sim T^3$, this production rate scales
linearly with temperature, whereas the Hubble rate is a quadratic
function of $T$.  It follows that for sub-MeV mass dark vectors, the
thermal production of $V$ is maximized at $T\sim m_e$.  However,
simple parametric estimates of this kind may require refinement due to
matter effects that alter the most naive picture. At finite
temperature $T$, the in-medium effects can be cast into a modification
of the mixing angle,
\begin{align}
 \label{eq:kappa-eff}
  \kappa_{T,L}^2 = \kappa^2 \times \frac{m_V^4}{|m_V^2 - \Pi_{T,L}|^2},
\end{align}
where $\Pi_{T,L}(\omega, |\vec q|, T)$ are the transverse (T) and
longitudinal (L) polarization functions of the photon in the isotropic
primordial plasma. They depend on photon energy $\omega $ and momentum
$|\vec q|$ and their temperature dependence is exposed by noting that
$\real \Pi_{T,L} \propto \omega_{\rm P}^2$ where $\omega_{\rm P}$ is
the plasma frequency; for the cases of interest
$\imag \Pi_{T,L}\ll \real \Pi_{T,L} $.

The consequences of these in-medium effects are two-fold. First, at high
temperatures, they suppress the mixing angle since
$\omega_{\rm P}^2\sim \alpha T^2$ (in the relativistic limit), thereby
diminishing contributions to thermal production for $T\gg m_V$.
Second, the presence of the medium allows the production to proceed
resonantly, whenever $\real \Pi_{T,L}(T_r,\omega) = m_V^2$ [process
(b) above].  Indeed, resonant conversion dominates the thermal dark
photon abundance for $m_V < 2 m_e$, but the
constraints from direct detection experiments rule out the possibility
of a thermal dark photon origin for $10\,\eV \lesssim m_V < 100\,\keV$
altogether. The values of $\kappa$ that are required for the correct
thermal relic abundance, estimated in
\cite{Pospelov:2008jk,Redondo:2008ec}, are  larger than the direct
detection bounds discussed here by several orders of magnitude.

Dark photon dark matter remains a possibility when the relic density
receives contributions from a vacuum condensate and/or from
inflationary perturbations, process~(c). The displacement of any
bosonic field from the minimum of its potential can be taken as an
initial condition, and during inflation any non-conformal scalar or
vector field receives an initial contribution to such displacements
scaling as $H_{\rm inf}/(2 \pi)$, where $H_{\rm inf}$ is the Hubble
scale during inflation.
Even in absence of initial misalignment, the inflationary production
of vector bosons can account for the observed dark matter density with
a spectrum of density perturbations that is commensurate with those
observed in the CMB~\cite{Graham:2015rva}.  While the production of
transverse modes is suppressed, longitudinal modes can be produced in
abundance~\cite{Graham:2015rva},
\begin{align}
 \label{eq:omg-inflation}
  \Omega_V  \sim 0.3 \sqrt{\frac{m_V}{1\,\keV}} 
  \left( \frac{H_{\rm inf}}{10^{12}\,\GeV} \right) .
\end{align}
For our mass range of interest the correct relic density would then be
attained with $H_{\rm inf}$ in the $10^{12}\,\GeV$ ballpark.

Undoubtedly, interactions between dark photons and the plasma are
present, and the evolution of any macroscopic occupation number of
vector particles is complicated by (resonant) dissipation
processes~\cite{Arias:2012az}.  For small enough couplings, these
processes may be made inefficient, and most of the vector particles
are preserved to form the present day DM. Equation
(\ref{eq:omg-inflation}) illustrates that---depending on the value
$H_{\rm inf}$---a successful cosmological model amenable to direct
detection phenomenology can always be found, and in the remainder of
this work we assume that $ \Omega_V h^2 = 0.12$, in accordance with
the CMB-inferred cosmological cold dark matter density. Consequently,
we also assume that the galactic dark matter is saturated by
$V$-particles, and neglect any effects from substructure. The latter
is a possibility when inflationary perturbations produce excess power
on very small scales~\cite{Graham:2015rva}, and which will make the
direct detection phenomenology ever more interesting. In this work, we
restrain ourselves to the smooth dark matter density and hence to the
time-independent part of the absorption signal.

\subsection{Stellar dark photon constraints}

In vacuum, this theory is exceedingly simple, as it corresponds to one
new vector particle of mass $m_V$ with a coupling $e\kappa$ to all
charged particles. Some of this simplicity disappears once the matter
effects for the SM photon become important, and the effective mixing
angle becomes suppressed. The subtleties of these calculations, taking  
proper account of the role of the longitudinal
modes of $V$, were fully accounted for only recently
\cite{An:2013yfc,An:2013yua,Redondo:2013lna,Graham:2014sha}. An understanding
of these effects is important because they determine the exclusion
limits set by the energy loss processes in the Sun, and other
well-understood stars \cite{Raffelt:1996wa}.  In the limit of small
$m_V$ (small compared to the typical plasma frequency in the central
region of the Sun), the energy loss into vector particles scales
as $\propto \kappa^2 m_V^2$, and is dominated by the production of
longitudinal modes \cite{An:2013yfc}. Although the resulting
constraints from energy loss processes turn out to be quite strong
in the $m_V \sim 100$ eV region, they weaken considerably for
very small $m_V$, opening a vast parameter space for a variety of
laboratory detection methods.

For $m_V > 10\,\eV$, dark matter experiments are sensitive enough to
compete with stellar energy loss bounds if dark photons contribute to a
significant fraction of the dark matter cosmological abundance. Here
we review the most important aspects of stellar emission for the
St\"uckelberg case, whereby we also update our previously derived
constraint on horizontal branch (HB) stars.

Ordinary photons inside a star can be assumed
to be in good local thermal equilibrium so that their distribution
function is time independent,
$\dot f_{\gamma}(\omega,T) = 0$.  This allows one to relate
photon production and absorption processes,
$d\Gamma_{\gamma}^{\rm prod}/d\omega dV = \omega |\vec q|/(2\pi^2)
e^{-\omega/T} \Gamma_{\gamma}^{\rm abs}$.
In analogy, for the production rate of on-shell dark photons one has,
\begin{align}
\label{eq:detailedbalance}
  \frac{d\Gamma^{\rm prod}_{T,L}}{d\omega dV}  = \kappa_{T,L}^2 
  \frac{\omega \sqrt{\omega^2 -m_V^2}}{2\pi^2} e^{-\omega/T} \Gamma^{\rm abs}_{\gamma, T,L},
\end{align}
where ${d\Gamma^{\rm prod}_{T,L}}/{d\omega dV}$ is the rate of
emission for a \mbox{spin-1} vector particle with mass $m_V$ and longitudinal
($L$) or transverse ($T$) polarization, while $\kappa_{T,L}^2 $ is defined in
(\ref{eq:kappa-eff}).  Inside active stars like our sun, the rate is
dominated by bremsstrahlung processes; for explicit
formulae see~\cite{An:2013yfc} and \cite{Redondo:2013lna}. The
expression (\ref{eq:detailedbalance}) is useful since the optical
theorem (at finite temperature) relates
$ \Gamma^{\rm abs}_{\gamma, T,L} = -\imag \Pi_{T,L}(\omega,\vec q)
/[\omega(1- e^{-\omega/T})]$.

Importantly, as alluded to above, emission can proceed resonantly
when $m_V^2 =\real \Pi_{T,L}$; see~(\ref{eq:kappa-eff}).  In the
emission of an on-shell dark photon,
$\real \Pi_L = \omega_P^2 m_V^2/\omega^2 $ and
$\real \Pi_T = \omega_P^2$, up to corrections of
${\cal O}(T/m_e)$. A resonance inside a star occurs when either
$\omega_P(r_{\rm res})^2 = \omega^2 $ (longitudinal) or
$\omega_P(r_{\rm res})^2 = m_V^2 $ (transverse). The emission
then proceeds from a spherical shell of radius $r_{{\rm res}}$ and
the rates become independent of the details of the emission process. One
may then integrate over the stellar profile by using the narrow width
approximation~\cite{An:2013yfc,Redondo:2013lna},
\begin{align}
\label{eq:resonant-rates}
  \frac{d\Gamma^{\rm prod}}{d\omega } &\simeq  \left(
  \frac{2 r^2 }{e^{\omega/T(r)} - 1}
  \frac{\sqrt{\omega^2 - m_V^2}} {|\partial \omega_P^2(r)/\partial r|} 
\right)_{r=r_{\rm res}} \nonumber\\
&\qquad \times \begin{cases}
    \kappa^2 m_V^2 \omega^2 & \text{longitudinal}, \\
    \kappa^2 m_V^4          & \text{transverse} , \\
  \end{cases} 
\end{align}
for each polarization of transverse $V$-bosons. 
This form nicely exhibits the different decoupling behavior with
respect to $m_V$. 
The bounds derived from stellar energy loss may qualitatively
be understood on noting that the typical plasma frequency at the center
of the star is given by,
\begin{align*}
 \text{Sun:} &\quad \omega_P(r=0)  \simeq 300\,\eV    ,\\
 \text{Horizontal Branch:}  & \quad\omega_P(r=0) \sim 2.6 \,\keV    ,\\
 \text{Red Giant:}  &\quad \omega_P(r=0) \sim 200\,\keV   ,
\end{align*}
and both longitudinal and transverse resonant emission stops
once $m_V> \omega_P(r=0)$.
In our numerical analysis, we employ the full expressions for
emission that also cover the case in which dark photons are emitted
off-resonance.

The shaded regions in Fig.~\ref{fig:summary} are a summary of
the astrophysical constraints on the mixing parameter $\kappa $ that are
independent of the relic density of dark photon dark matter. The thin solid
(dotted) gray lines show the constraints that are based solely on the
emission of transverse (longitudinal) modes.

For the sun, the limit on the anomalous energy loss rate is identical
to the one in previous work~\cite{An:2013yfc, Redondo:2013lna}. As a criterion
we require that the luminosity in dark photons cannot exceed 10\% of
the solar luminosity, $L_{\odot} = 3.83\times 10^{26}$~W. The limit is
derived from observations of the $^8$B neutrino flux; for details we
refer the reader to the above references.

For Horizontal Branch (HB) stars, we update our own previously derived
limit as follows (a similar limit has already been presented
in~\cite{Redondo:2013lna}): as an HB representative, we consider a
$0.8 M_{\odot} $ solar mass star with stellar profiles as shown in
in~\cite{Raffelt:1996wa,Dearborn:1989he}. The energy loss is then
limited to 10\% of the HB's luminosity~\cite{Raffelt:1996wa}, for
which we take $L_{\rm HB} = 60 L_{\odot}$~\cite{Dearborn:1989he}.  The
transverse modes dominate the energy loss in HB stars. Since the
corresponding resonant emission originates from one shell
$r_{\rm res, T}$ for all energies, the derived constraint is sensitive
to the stellar density profile in the resonance region
$m_V < \omega_P(r=0)\simeq 2.6\,\keV$. For example, the kink visible
in the thin gray line at $m_V\sim 150\,\eV$ originates from entering
the He-burning shell.  Our result is in qualitative agreement
with~\cite{Redondo:2013lna}; quantitative differences may be assigned
to our use of full emission rate expressions [rather then
(\ref{eq:resonant-rates})] that are integrated using Monte Carlo
methods over the assumed stellar profile. In either case, such bounds
are---by construction---only representative in nature and a detailed
comparison of the derived limits will not yield much further insight.

Finally, the constraint that can be derived from Red Giant (RG) stars
extends sensitivity to larger $m_V$. We require a dark photon
luminosity that is less then $10\,{\rm erg}/{\rm g}/\sec$ originating
from the degenerate He core with $\rho \sim 10^6\,{\rm g}/\cm^3 $,
$T\simeq 8.6\,\keV$~\cite{Raffelt:1996wa}. Longitudinal emission
dominates until transverse emission becomes resonant at
$m_V = \omega_P({\rm core}) \sim 20\,\keV$.  Here we note that there
is room for improvement when deriving the limit from RG stars. For
example, recent high-precision photometry for the Galactic globular
cluster M5 has allowed the authors of~\cite{2013PhRvL.111w1301V} to
derive constraints on axion-electric couplings and neutrino dipole
moments that are based on the observed brightness of the tip of the RG
branch. In conjunction with an actual stellar model, however, the
better observations do not yield a drastic improvement of limits, as
there appears to be a slight preference for extra
cooling~\cite{2013PhRvL.111w1301V}. Albeit such hints to new physics
are tantalizing, we in turn expect only mild changes to our
representative RG constraint when a detailed stellar model is employed
and/or better observational data is used; we leave such study for the
future.

\subsection{Constraints from $V\to 3\gamma$ decay}
\label{sec:constraints-from-vto}

Next we consider constraints imposed by energy injection from
$\gamma$-rays originating from $V\to 3\gamma$ decays below the
$e^+e^-$ threshold, for which the one-photon inclusive differential
rate was computed in~\cite{Pospelov:2008jk}. It reads,
\begin{align}
\label{eq:threegamma}
  \frac{d\Gamma}{dx} = \frac{\kappa^2\alpha^4}{2^73^75^3\pi^3}
 \frac{m_V^9}{m_e^8} x^3 \left[ 1715 - 3105x + \frac{2919}{2} x^2 \right],
\end{align}
where $x = 2 E_{\gamma}/m_V$ with $0\leq x \leq 1$; the total decay
width is obtained by integration,
$\Gamma_{V\to 3\gamma} = \int_0^1 dx\, d\Gamma/dx$, and it sets the
lifetime of dark photons for $m_V < 2m_e$. 

A limit from observations of the diffuse $\gamma$-ray background was
estimated in~\cite{Redondo:2008ec} by translating the results for
monochromatic photon injection obtained in \cite{Yuksel:2007dr} and
assuming a photon injection energy of $m_V/3$. Here we re-consider
this limit and base it on the actual shape of the inclusive one-photon
decay spectrum~(\ref{eq:threegamma}). Let $dN/dE_{\gamma}$ denote the
differential spectrum such that
$\int dE_{\gamma} \, dN_{\gamma}/dE_{\gamma} = 3$. It follows that
$E_{\gamma} (dN/dE_{\gamma}) = 3 \Gamma_{V\to 3\gamma}^{-1} x
(d\Gamma/dx )$.

There are then two contributions to the diffuse photon background from
$V\to 3\gamma$ decays. For the flux from the dark matter density at
cosmological distances we find,
\begin{align}
  E_\gamma \frac{d\phi_{\rm eg}}{dE_{\gamma}} =   \frac{\Omega_V \rho_c \Gamma_{V\to3\gamma}}{4\pi m_V} 
  \int_0^{z_f} dz\, \frac{ E_{\gamma}}{H(z) } \frac{ dN[(1+z)E_{\gamma}]}{ dE_{\gamma}} ,
\end{align}
where we have made the assumption that most of the dark matter has not yet
decayed today, $\Gamma_V\ll H_0$, with $H_0$ being the present day
Hubble rate. $H(z)$ is the Hubble rate at redshift $z$ and we cut off
the integral at the (blueshifted) kinematic boundary,
$ z_f = m_V/(2E_{\gamma}) - 1$, or, for $E_{\gamma}\to 0 $, at some
maximal redshift that is numerically inconsequential.
In turn, the galactic diffuse flux is given by,
\begin{align}
\label{eq:Jgal}
  E_\gamma \frac{d\phi_{\rm gal}}{dE_{\gamma}} =   \frac{ \Gamma_{V\to 3 \gamma}   }{4\pi m_V} 
  E_{\gamma} \frac{dN}{dE_{\gamma}}  \rho_{\rm sol}  R_{\rm sol} \mathcal{J}, 
\end{align}
where $\mathcal{J(\psi)}$ is the
$\rho_{\rm sol} R_{\rm sol}$--normalized line-of-sight integral at an
angle $\psi$ from the galactic center;
$\rho_{\rm sol} \simeq 0.3\,\GeV/\cm^3 $ is the dark matter density at
the sun's position, $R_{\rm sol} \simeq 8.3\,\rm kpc $ away from the
galactic center. For estimating the diffuse photon contribution,
taking $\psi = \pi$ or $\pi/2$ yields
$\mathcal{J} \simeq 1$ or $1.6$ for a NFW or an Einasto dark matter
density profile, and we take $\mathcal{J}=1$ as fiducial value in
(\ref{eq:Jgal}).%

\begin{figure}
\centering
\includegraphics[width=\columnwidth]{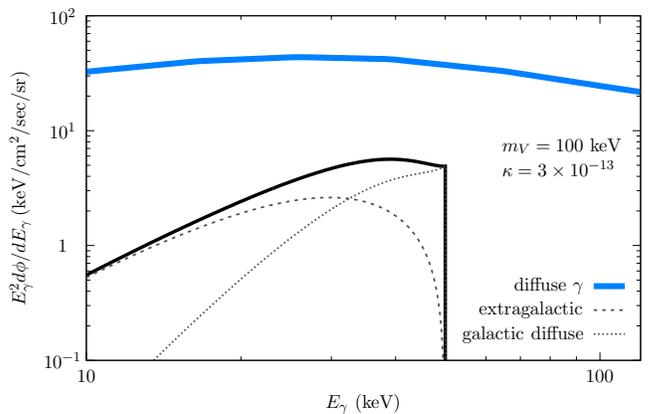}%
\caption{\footnotesize Representative diffuse gamma ray bolometric
  flux (thick solid top line) together with computed extragalactic
  (galactic) photon fluxes depicted by the dashed (dotted) line from
  $V\to 3\gamma$ decay. We constrain the sum of these fluxes (solid
  line) to not exceed the observed one.}
\label{fig:diffuse}
\end{figure}

Figure~\ref{fig:diffuse} depicts the representative diffuse gamma ray
flux of photons (thick solid line) as taken
from~\cite{Yuksel:2007dr}. The extragalactic and diffuse galactic
fluxes originating from dark photon decay with $m_V = 100\,\keV$ and
$\kappa = 3\times 10^{-13}$ are respectively shown by the dashed and
dotted lines.  We constrain the flux contribution from dark photon
decay by requiring that their sum (solid line) does not exceed 100\%
(10\%) of the observed flux. The ensuing limits in the $(m_V,\kappa)$
parameter space are shown in Fig.~\ref{fig:summary} and they constrain
the region $m_V > 100\,\keV $. While the derived limit represents a
conceptual improvement because use of the differential photon spectrum
has been made, quantitatively, the strength of the limits is
comparable to the previous estimate~\cite{Redondo:2008ec}.

The final constraint discussed in this section is due to precise measurements of the cosmic
microwave background (CMB) radiation, and its sensitivity to DM decay. 
Specifically, $V\to 3\gamma$ decays at
redshift $O(1000)$ alter the ionization history, raising TE and EE
amplitudes on large scales, and damping TT temperature fluctuations on
small scales. An energy density of
\begin{align}
  \frac{dE}{dV dt} =  3 \zeta m_p \Gamma_{V\to 3 \gamma} e^{-\Gamma_{V\to 3\gamma}t} 
\end{align}
is injected into the plasma per unit time where
$\zeta = (f/3) \Omega_V/\Omega_b$ is related to the injected energy
per baryon, which is equal to $3\zeta m_p$; $m_p$ is the mass of the proton and $f$
denotes the overall efficiency with which the plasma is heated and
ionized. In the case at hand $f=1$. In \cite{Fradette:2014sza} limits
on the combination $(\zeta,\Gamma_V)$ were derived for decaying heavy
dark photons with $m_V>2m_e$, utilizing the Planck 2013 and WMAP
polarization data. (For earlier analyses, see {\em e.g.}
\cite{Chen:2003gz,Cline:2013fm}.)  For lifetimes significantly longer
than the cosmic time of recombination, the limit amounts to
$\zeta \Gamma_V\lesssim 6\times 10^{-17}\,\eV/\sec$ or
$\tau_V \gtrsim 10^{26}\,\sec$. We show this constraint in
Fig.~\ref{fig:summary} and it is very comparable in sensitivity to the
one derived from diffuse $\gamma$-ray lines.

\section{DM absorption signals in direct detection experiments}
\label{sec:dm-absorpt-sign}

\subsection{Dark vector-induced ionization}

If the energy of dark vectors is above the photoelectric threshold
$E_V \geq E_{\rm th}$, atomic ionization becomes viable, for example
in Xenon:
\begin{equation}
{\rm Xe~I} + V \to {\rm Xe~II} +e^-;~~{\rm Xe~I} + V \to {\rm Xe~III} +2e^-;...
\label{ion} 
\end{equation}
Here I's are used according to the usual atomic notation, and Xe I represents
the neutral Xenon atom which is most relevant for our discussion.
Most of the DM is cold and non-relativistic, so that $E_V = m_V$ with
good accuracy. The astrophysics bounds, on the other hand, are often
derived in the regime $E_V \gg m_V$. We will address the
$E_V \simeq m_V$ case first, where the distinction between L and T
modes all but disappears.

When $m_V \geq E_{\rm th}=12.13$ eV, matter effects are not very
important, and the problem reduces to the absorption of a massive
nonrelativistic particle with $e\kappa$ coupling to electrons. The
difference with the absorption of a photon with $\omega = m_V$ amounts
to the following: the photon carries momentum $|\vec{q}| =\omega$,
whereas the nonrelativistic dark vector carries a negligibly small
momentum, $|\vec{q}|= m_V  v_{\rm DM} \sim O(10^{-3})\omega$
where $v_{\rm DM}$ is the dark photon velocity. Fortunately, this
difference has little effect on the absorption rate for the following
reason. Both the photon wavelength and the DM Compton wavelength are
much larger than the linear dimension of the atom, allowing for a
multipole expansion in the interaction Hamiltonian,
$(\vec{p}_e \vec{\epsilon})\exp(i\vec{q}\vec{r}_e) \simeq (\vec{p}_e
\vec{\epsilon})\times(1 + i \vec{q}\vec{r}_e +...)$,
where $\vec{\epsilon}$ is the (dark) photon polarization. The first
term corresponds to the E1 transition that dominates over other
multipole contributions, making the matix elements for absorption of
`normal' and dark photons approximately equal.  Accounting for the
differences in flux, and averaging over polarization, gives the
relation between the absorption cross sections~\cite{Pospelov:2008jk}
\begin{equation}
\sigma_V(E_V=m_V)v_{V} \simeq \kappa^2\sigma_\gamma(\omega=m_V)c,
\label{relation}
\end{equation}
where $v_{V}$ is the velocity of the incoming DM particle.  This
relation is not exact and receives corrections of order
$O(\omega^2 r_{\rm at} ^2)$ where $r_{\rm at} $ is the size of
corresponding electronic shell participating in the ionization
process. Near ionization thresholds this factor varies from
$\sim \alpha^2$ for outer shells to $\sim Z^2 \alpha^2$ for inner
shells.  We deem this accuracy to be sufficient, and point out that
further improvements can be achieved by directly calculating the
absorption cross section for dark photons using the tools of atomic
theory. (Analogous calculations have already been performed for the
case of axion-like DM \cite{Dzuba:2010cw}.)

Relation (\ref{relation}) is nearly independent of the DM velocity,
and results in complete insensitivity of the DM absorption signal to the
(possibly) intricate DM velocity distribution in the
galactic halo; this is in stark contrast to the case of WIMP elastic
scattering.  The resulting absorption rate is given by
\begin{equation}
{\rm Rate~per~atom} \simeq \frac{\rho_{\rm DM}}{m_Vc^2} \times \kappa^2\sigma_\gamma(\omega=m_V)c,
\label{rate}
\end{equation}
where $\rho_{\rm DM}$ is the local galactic DM energy density, and
factors of $c$ are restored for completeness.

The above formulae are sufficiently accurate provided all medium
effects can be ignored.  In general, however, the process of
absorption of a dark photon must also account for the modification of
$V-\gamma$ kinetic mixing due to in-medium dispersion effects.  While
the absorption of $m_V \gg E_{\rm th}$ particles cannot be affected
significantly, close to the lowest theshold such effects can be
important.
To account for in-medium effects, we follow our original derivation
in~\cite{An:2013yua}.  The matrix element for photon absorption
$q + p_i \to p_f$ with photon four momentum $q=(\omega, \vec q)$ and
transverse ($T$) or longitudinal ($L$) polarization vectors
$\epsilon_{\mu}^{T,L}$ is given by,
\begin{align}
  \mathcal{M}_{i \to f + V_{T,L}} = - \frac{e \kappa m_V^2}{m_V^2
    -\Pi_{T,L}(q)} \langle p_f |  J^{\mu}_{em} (0) | p_i \rangle
  \varepsilon_{\mu}^{T,L}(q) .
\end{align}
Squaring the matrix element and summing over final states $f$, one
obtains the  absorption rate of L or T photons,
\begin{align}
\label{eq:absorptionStueck}
  \Gamma_{T,L} & = \frac{1}{2\omega} (2\pi)^4 \delta^{(4)} (q + p_i - p_f) e^2
  \kappa_{T,L}^2 \varepsilon_{\mu}^{*} \varepsilon_{\nu} \nonumber\\
  & \qquad\qquad \times \sum_f
  \langle p_i | J^{\mu}_{em} (0) | p_f \rangle \langle p_f |
  J^{\nu}_{em} (0) | p_i \rangle  \\
  & = \frac{e^2}{2\omega} \int d^4x\, e^{i q\cdot x} \kappa_{T,L}^2
  \varepsilon_{\mu}^{*} \varepsilon_{\nu} \langle p_i | [ J^{\mu}_{em} (x) ,
  J^{\nu}_{em} (0) ] | p_i \rangle ,
\end{align}
where the in-medium effective $V-\gamma$ mixing angle is given in
(\ref{eq:kappa-eff}).
The polarization functions $\Pi_{T,L}$ are obtained from the in-medium
polarization tensor $\Pi^{\mu\nu}$,
\begin{align}
\label{eq:poltensor}
  \Pi^{\mu\nu}(q) &= i e^2 \int d^4x\, e^{i q\cdot x} \langle 0 |
  T J^{\mu}_{em} (x) J^{\nu}_{em} (0) | 0 \rangle \nonumber \\
  & = - \Pi_T
  \sum_{i=1,2} \varepsilon_i^{T\mu} \varepsilon_i^{T\nu} -
  \Pi_L\varepsilon^{L\mu} \varepsilon^{L\nu} .
\end{align}
Noting that
\begin{align}
  &\int d^4x\,e^{i q\cdot x}  \langle 0 | [ J^{\mu}_{em} (x) , J^{\nu}_{em} (0) ] | 0
  \rangle \nonumber\\
  &\qquad = 2 \imag{ \left[ i \int d^4x\,e^{i q\cdot x}  \langle 0 | T J^{\mu}_{em}
      (x) J^{\nu}_{em} (0) | 0 \rangle \right] },
\end{align}
we can express the absorption rate in the lab-frame of the
detector (\ref{eq:absorptionStueck}) as follows,
\begin{align}
\label{eq:Gamma}
\Gamma_{T,L} & = - \frac{\kappa_{T,L}^2 \imag{\Pi_{T,L}}}{\omega} . 
\end{align}

\begin{figure*}
\centering
\includegraphics[width=0.40\textwidth]{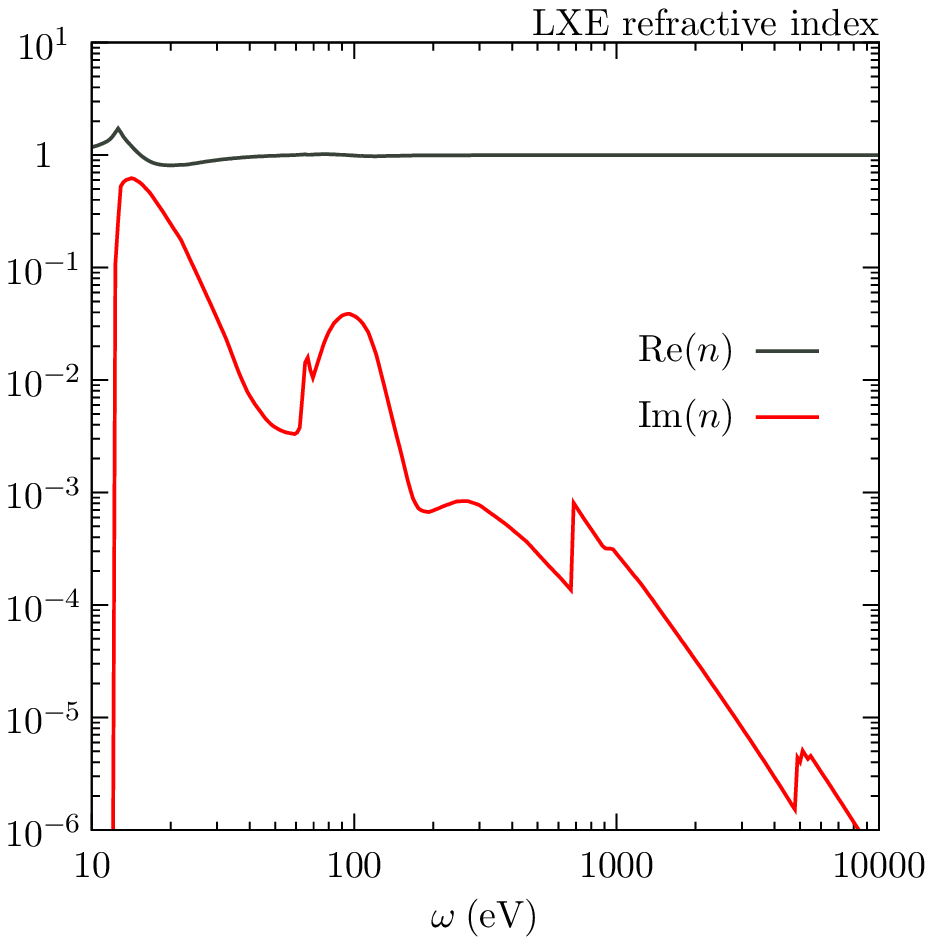}%
\includegraphics[width=0.45\textwidth]{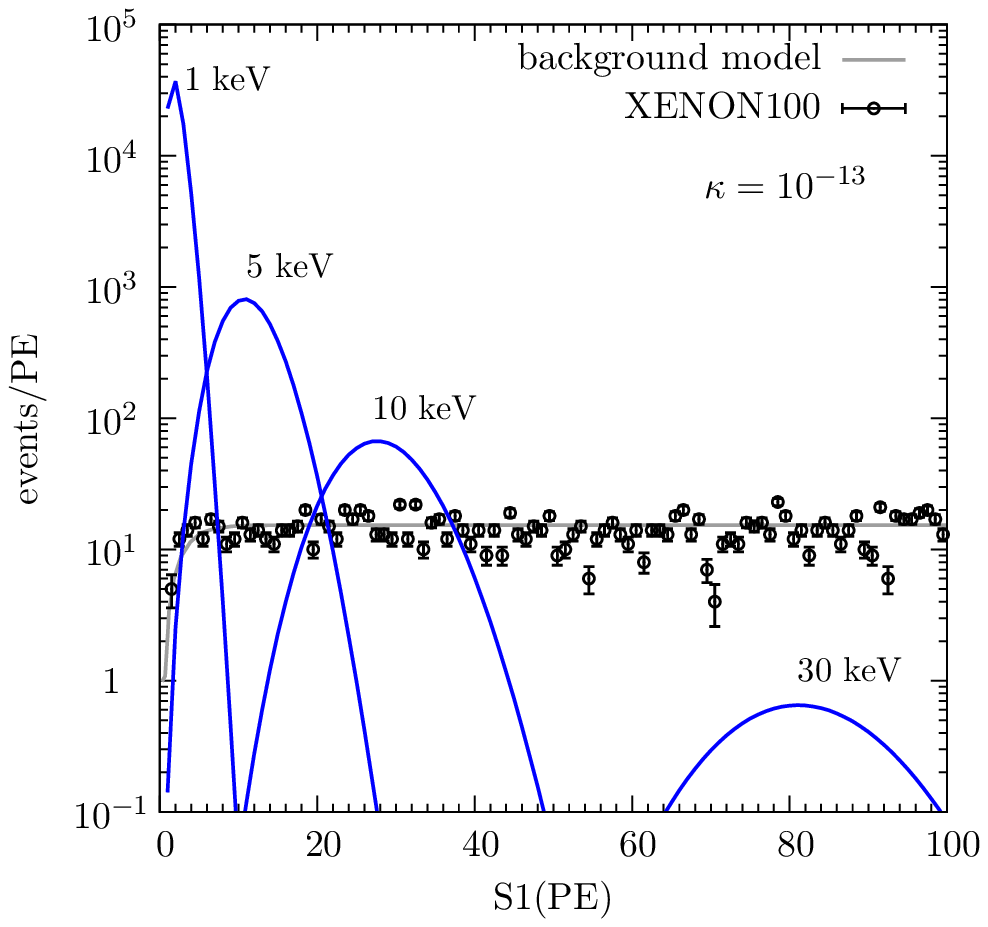}%
\caption{\footnotesize \textit{Left:} Real and imaginary parts of the
  liquid xenon refractive index computed from tabulated atomic
  scattering factors and using the Kronig-Kramers relation. Note that
  the maximum of the $\imag (n)$ function corresponds to the
  photoelectric cross section
  $\sigma_\gamma \sim 6\times 10^{-17}{\rm cm}^2$.  \textit{Right:}
  Simulated events in `xenon-units' of photo-electrons (PE) for
  various dark photon masses as labeled. Also shown are the reported
  event counts and the background model as taken
  from~\cite{Aprile:2014eoa}.}
\label{fig:xs}
\end{figure*}

This particular form is suitable for calculation, as we can relate $\Pi_{T,L}$ to
tabulated optical properties of the material. For an isotropic and
non-magnetic medium,
\begin{align}
  \Pi_{L} = (\omega^2 - \vec q^2) (1- n_{\mathrm{refr}}^2),\quad 
   \Pi_T = \omega^2 (1- n_{\mathrm{refr}}^2),
\end{align}
where $n_{\rm refr}$ is the (complex) index of
refraction for electromagentism. When $|\vec{q}|\ll \omega$, $\Pi_L=\Pi_T\simeq \Pi$, 
and all formulae for the absorption of L and T modes become idential, 
as expected. 

As the final step, we obtain $ n_{\mathrm{refr}} $ from its relation to
the forward scattering amplitude $f(0) = f_1 + i f_2$, where the atomic
scattering factors $f_{1,2}$ are tabulated \textit{e.g.}
in~\cite{Henke:1993eda}. Close to the ionization threshold we make use
of the Kramers-Kronig dispersion relations to relate $f_1$ and $f_2$
in estimating $n_{\rm refr}$. Alternatively, one may establish an
integral equation relating the real and imaginary parts of
$\varepsilon$; see~\cite{An:2013yua}.

When $m_V^2 \gg \Pi$, $\kappa_{L(T)} \simeq \kappa$, and the in-medium
modification of absorption can be negelected. In that case the
absorption rate per DM particle is
\begin{equation}
\Gamma \simeq \kappa^2 \omega \times  {\rm Im }~ n_{\mathrm{refr}}^2 = \kappa^2 \sigma_{\gamma}\times \left(\frac{N_{\rm at}}{V} \right),
\end{equation}
leading to the same formula for the absorption rate per atom as
before, Eq.~(\ref{rate}).

\subsection{XENON10}
\label{sec:xenon10-1}

The XENON10 data set from 2011 exemplifies the power of
ionization-sensitive experiments when it comes to very low-energy
absorption-type processes. With an ionization threshold of
$\sim 12\,\eV$, the absorption of a 300~eV dark photon already yields
about 25 electrons, and the relatively small exposure of 15~kg-days is
still sufficient to provide the best limits on dark photons originating
from the solar interior~\cite{An:2013yua}. The same type of signature 
is used to provide important contraints on WIMP-electron scattering \cite{Essig:2011nj,Essig:2012yx}. 

Despite significant uncertainties in electron yield, energy
calibration, and few-electron backgrounds, we would like to emphasize the
fact that \textit{robust} and \textit{conservative} limits can be derived which are 
independent of the above systematics. The procedure is straightforward, and 
follows the one already outlined in~\cite{An:2013yua}.
First, we count all ionization events (246) with up to 80 ionization
electrons, or, equivalently, within 20~keV of equivalent nuclear
recoil.  If we do not attempt to subtract backgrounds (which is conservative), this
implies a $90\%$ C.L.~upper limit of less than $19.3$ dark photon
absorptions per kg per day---irrespective of how many electrons are
ultimately produced (as long as the number is less than 80.)  From
that integral limit we derive the ensuing XENON10 dark photon dark
matter constraint shown in Fig.~\ref{fig:summary}. Remarkably, we
observe that for $ 12\,\eV \lesssim m_V\lesssim 200\,\eV $ the new
limit is stronger than the previously derived solar energy loss
constraint.

\subsection{XENON100}
\label{sec:xenon100}

The XENON100 collaboration has performed a low-threshold search using
the scintillation signal S1 with an exposure of 224.6 live days and an
active target mass of 34~kg liquid xenon \cite{Aprile:2014eoa}. A very low background rate
of $\sim 5\times 10^{-3}/\kg/\day/\keV $ has been achieved through
a combination of xenon purification, usage of ultra-low radioactivity
materials, and through self-shielding by volume fiducialization.
In addition, with energy deposition in the keV range and above, the XENON100 
experiment provides a sufficient energy resolution, allowing for 
mass reconstruction of a potential DM absorption signal. 

We derive the signal in the XENON100 detector as follows. For the dark
photon dark matter the kinetic energy is negligible with respect to
its rest energy since $(v/c)^2 \sim 10^{-6}$. Therefore, a
mono-energetic peak at the dark photon mass is expected in the
spectrum. To derive the constraint, we first convert the absorbed
energy $m_V$ into the number of photo-electrons (PE) using Fig.~2 of
Ref.~\cite{Aprile:2014eoa}. This may result in a 10\% uncertainty due
to the corrections from binding energies of electrons at various
energy levels as shown in Fig. 1 of Ref.~\cite{Szydagis:2013sih}. We
take into account the Poissonian nature of the process, and include
the detector's acceptance as a function of S1, shown in Fig.~1 of
Ref.~\cite{Aprile:2014eoa}. The resulting S1 spectrum for various dark
photon masses together with the reported data is shown in
Fig.~\ref{fig:xs}.

A likelihood analysis is used to constrain the kinetic mixing
$\kappa$. The likelihood function is defined as
\begin{equation}
L(\kappa, m_V) = \prod_{i\geq3} {\rm Poiss}(N^{(i)}|N_s^{(i)}(\kappa, m_V)+N_b^{(i)}) \ ,
\end{equation}
where $i$ labels the bin number (which equals the number of S1 for
each event) as shown in Fig.~6 of Ref.~\cite{Aprile:2014eoa},
$N_b^{(i)}$ and $N^{(i)}$ are the background and number of observed
events as presented in Ref.~\cite{Aprile:2014eoa}. Following the
latter experimental work, we apply a cut S1$\geq3$. Here we neglect
the contribution from the uncertainty of $n^{exp}$ to the likelihood
function, since from Fig.~2 of Ref.~\cite{Aprile:2014eoa} one can see
that after we apply the S1$\geq3$ cut, its influence on the limit of
$\kappa$ is less than 10\%. A standard likelihood analysis then yields
the resulting $2\sigma$  limit on $\kappa$ as a function of
$m_V$. It is shown as the black dashed curve in Fig.~\ref{fig:summary}.
Again, we find the direct detection constraints to be very competitive
with astrophysical bounds.

\section{Discussion and Conclusion}
\label{sec:disc-concl}

With an array of direct detection experiments now searching for
signatures of elastic nuclear recoil of WIMPs on nuclei, and with
sensitivity levels marching towards the neutrino background, it is
important to keep in mind that other dark matter scenarios can also be
sensitively probed with this technology. In particular, the exquisite
sensitivity to ionization signatures at various experiments allows
stringent constraints to be placed on generic models of
super-weakly-interacting dark matter. In this paper, we have studied
the sensitivity to the minimal model of dark photon dark matter, and
obtained limits (summarized in Fig.~\ref{fig:summary}) that exceed
those from stellar physics over a significant mass range.

The sensitivity of current direct detection experiments already
excludes dark photon dark matter with a thermally generated abundance.
This is not a problem for the model, as the DM abundance may be
determined by non-thermal mechanisms. For example, perturbations
during inflation may create the required relic
abundance~\cite{Graham:2015rva}, and further constraints on such
models may be achieved if an upper bound on $H_{\rm inf}$ were to be
established by experiments probing the CMB.

Dark photon dark matter has certain advantages over
axion-like-particle dark matter with respect to direct detection. The
absence of the dark photon decay to two photons removes the constraint
from monochromatic X-ray lines. This latter signature usually provides a more
stringent constraint on axion-like keV-scale DM than direct detection. Furthermore, 
the cross section for dark photons is significantly
enhanced for small masses, relative to the cross section for
absorption of axion-like particles.

The analysis presented in this paper addresses the model of a very
light dark photon field, that is particularly simple and
well-motivated. In addition, one could construct a whole family of
`simplified' models of very light dark matter, with observational
consequences for direct detection \cite{Pospelov:2008jk}. The most
relevant of these would involve couplings to electrons, and one could
consider DM of different spin and parity:
\begin{align}
\label{variety}
&{\rm (pseudo)scalar}~~~g_S S \bar \psi \psi,~~ g_P P \bar \psi \gamma_5 \psi, \nonumber \\
&{\rm (pseudo)vector}~~~g_V V_\mu \bar \psi \gamma_\mu \psi, ~~ g_A {\cal A}_\mu \bar \psi \gamma_\mu \gamma_5 \psi, \\
&{\rm tensor}~~~g_T T_{\mu\nu} \bar \psi \sigma_{\mu\nu} \psi, ~\cdots \nonumber
\end{align}
Here $\psi$ stands for the electron field, $g_i$ parametrizes the
dimensionless couplings, and $V,{\cal A},S,P,T...$ are the fields of
metastable but very long lived DM. The case considered in this paper
corresponds to $g_V = e\kappa$, and the light mass $m_V$ is protected by
gauge invariance.  However, even cases where the mass of DM is not protected by any symmetry are of
interest, and can be considered within effective (or simplified)
models. In this case, loop processes tend to induce a finite mass
correction, which is at most
$\Delta m_{{\rm DM}i} \sim g_i \Lambda_{\rm UV}$. With the cutoff
$\Lambda_{\rm UV}$ at a TeV, it is natural to expect that, for a
DM mass of $\sim$ 100~eV for example, one should have $g_i< 10^{-10}$. As
demonstrated by the analysis in this paper, DM experiments can
probe well into this naturalness-inspired regime, and set meaningful constraints
on many variations of light DM models.

Finally, we would like to emphasize that further progress can be
achieved through the analysis of `ionization-only' signatures.  For
example, in noble gas- and liquid-based detectors one can improve the
bounds for $E<\keV$ by accounting for multiple ionization electrons
(see Ref.~\cite{Essig:2012yx}). The ionization of Xe atoms from the lowest
electronic shells is likely accompanied by Auger processes,
which generate further photo-electrons, and the corresponding bounds
can be tightened.  Analysis of these complicated processes may require
additional input from atomic physics.

\subsection*{Acknowledgements}

We would like to thank Fei Gao, Liang Dai, and Jeremy Mardon for helpful
discussions. HA is supported by the Walter Burke Institute at Caltech
and by DOE Grant DE-SC0011632. The work of MP and AR is supported in
part by NSERC, Canada, and research at the Perimeter Institute is
supported in part by the Government of Canada through NSERC and by the
Province of Ontario through MEDT. JP is supported by the New Frontiers
program of the Austrian Academy of Sciences.

\end{document}